\documentclass{aa}  

\usepackage{natbib}
\usepackage{graphicx}
\usepackage[colorlinks=true, citecolor=blue, linkcolor=black, urlcolor=black]{hyperref}
\usepackage{txfonts}
\usepackage{lipsum}
\usepackage{subcaption}
\usepackage{lscape}
\usepackage{placeins}

\newcommand{\cA}{\mathcal{A}}
\newcommand{\cB}{\mathcal{B}}

\newcommand{\cI}{\mathcal{I}}

\newcommand{\cZ}{\mathcal{Z}}

\newcommand{\vv}[1]{\mathbf{#1}}

\begin{document}

   \title{Fine-tuning of light-time effect in triple systems}

   \author{David Vokrouhlick\'y}
   
   \institute{Institute of Astronomy, Charles University, V Hole\v{s}ovi\v{c}k\'{a}ch 2,
              CZ-18000 Prague 8, Czech Republic\\
              \email{vokrouhl@cesnet.cz}}
              
   \date{Received \today}

  \abstract
   {The sequence of eclipses of binary stars is subject to inequalities for various reasons.
    The presence of a third component in the system causes periodic motion of the binary's
    center of mass along the line of sight of an observer. The finite value of the light
    velocity implies that the epochs of eclipses periodically advance and delay with respect
    to the exact orbital period of the binary, a phenomenon termed the light-time effect (LITE).}
   {We aim to refine two aspects of the mathematical treatment of LITE. First, we provide
    both generalized and more accurate analytic formulation describing the light-travel 
    time in the binary system itself presented in previous works. Second, we 
    analytically estimate the so far neglected coupling of LITE with the dynamical 
    interaction of the binary orbit with the motion of the third star.}
   {Our principal results are given in a simple analytical form, which is suitable for
    the analysis of photometric observations that require minimization over a multidimensional
    parameter space of the triple system.}
   {The leading correction to the traditional formulation of LITE due to the light-travel time
    in the binary system may be detectable for triple systems with a period ratio of
    $P_2/P_1\lesssim 20$, for which accurate photometric observations are available. On the
    other hand, the correction due to the dynamical coupling of the two orbits with $P_2$
    periodicity is small, but may become relevant in the future.}
   {}

   \maketitle

   \nolinenumbers 

\section{Introduction}
Interesting scientific problems usually have a long and somewhat complex history. That is also 
the case of variations in eclipse epochs of binary stars embodied in systems of a higher 
hierarchy attributed to the finite speed of light. As the center of mass (COM) of the binary 
periodically moves along the line of sight to the observer, due to its revolution about the 
global COM of the whole system, epochs of eclipses are either delayed or advanced; this phenomenon 
is often termed the light-time effect (LITE; though some authors prefer the light-travel-time 
effect; LTTE). However, even in the simplest of such situations, triple-star systems, the 
variations in eclipse epochs may arise for several other reasons, including mutual interactions 
of the components in the binary or gravitational perturbations of the three stars in the 
system. Separation of the role of these different phenomena proved tricky and required only a 
sufficiently long series of good-quality photometric and spectroscopic observations.

The origin of the idea concerning the role of LITE in describing inequalities of eclipses of 
variable stars belongs to \citet{chan1888,chan1892}, who analyzed century-long observations of 
Algol ($\beta$~Persei). Although pleased with the concept of the phenomenon, \citet{tiss1895} 
pointed out another solution of the data, which was further supported by analysis of the 
spectroscopic observations by \citet{cur1908}.%
\footnote{Modern data confirm the LITE contribution to Algol's eclipse timing variations
 complemented by a complex mixture of contributions of dynamical origin \citep[e.g.,][]{fri1970}.}
Nevertheless, the seed remained, and, on the suggestion of Ejnar Hertzsprung \citep[see][]{hertz1922}, 
the first formal mathematical analysis of LITE was developed by \citet{wol1922}; the method was 
later extended by \citet{irw1952}. However, progress in terms of the photometric observations was 
rather slow. Their available accuracy implied that LITE may be detected for systems of a 
sufficiently long period, $P_2,$ of the outer orbit, which required long series of data and 
supportive spectroscopic evidence of the third star in the system. Reviews by \citet{fch1973} 
and \citet{mayer1990} concluded that there were only a handful of stellar systems with an 
unambiguous detection of LITE.

Things have dramatically changed in the last two decades or so, mainly due to very accurate 
space-born photometric surveys of the Kepler, CoRoT, and TESS missions. A breathtaking set 
of multiple stellar systems of all possible geometries and mass regimes was discovered and 
accurately characterized
\citep[e.g.,][]{rappa2013,conroy2014,borko2015,borko2016,borko2022}. At the same time, the 
theory of dynamical interactions in triple systems, including their implications for eclipse 
timing variations, has seen significant improvements 
\citep[e.g.,][]{borko2003,borko2011,borko2015,borko2022}. As a result, the portfolio of precise 
data and accurate tools for their analysis has significantly increased. 

In contrast to these huge advances in theoretical description of the eclipse inequalities due 
to physical interaction between stars in triple systems, the theory of LITE did not see 
much development in the context of stellar systems for a long time following the Irwin 
formulation. New ideas came with the first detections of exoplanets using transits of their 
hosting stars, and especially the first successful observations of inequalities in their 
periodicity \citep[transit timing variations; e.g.,][]{fab2010}. Since the same space-born 
photometric surveys furnished the data for transiting exoplanets and eclipsing binaries, 
stellar astronomers brought the new findings on detailed modeling of light-travel effects 
to their field as well. For example, \citet{kap2010} considered the asymmetric shift in 
accurately measured primary and secondary eclipses of white dwarf systems due to the finite 
speed of light to constrain the stellar masses. The most recent remarkable advancement in 
LITE modeling was presented by \citet{conroy2018}. Apart from the traditional LITE due to 
\emph{radial} motion of the binary with respect to the observer, these authors pointed out 
that asymmetry in the \emph{transverse} motion of the two stars in the binary should cause 
an additional shift in the observed eclipses. They developed a conceptually general
formulation of the effect and applied it to several relevant situations, including 
hierarchical triple systems. Arguing that the effect is maximal for coplanar systems, 
they focused their analysis on this case. Studying the LITE correction of the same physical
essence as \citet{conroy2018}, namely due to the corrected positions of stars
at the center of eclipse caused by the finite travel time of light in the binary
system, we used a different approach that allowed us to obtain more general and
accurate results. In particular, we provide an analytical formulation for a general
orbital architecture of triple systems. This helped us to find that
the same maximal effect discussed in the coplanar case also occurs in various
non-coplanar systems.

In what follows, we consider two refinements of LITE analytical modeling. After 
briefly introducing notation and reviewing the traditional analysis of LITE in 
Sect.~\ref{liteold}, in Sect.~\ref{litenew1} we provide a detailed analysis of the role 
of the finite travel time of light in the binary for generally non-coplanar systems. We
restrict ourselves to the case of small eccentricity of the binary, but allow an 
arbitrary eccentricity of the orbit of the third star in the system. 
In Sect.~\ref{litenew2}, we analyse LITE shift due to mutual gravitational perturbation of
the binary orbit and the orbit of the third star. We provide an analytic estimate of the
principal effect with periodicity of the outer orbit in the triple. Since we find this LITE
correction small, we restricted ourselves to the coplanar geometry of the triple system. 
Section~\ref{concl} concludes our findings and provides an outlook for future developments.

\section{Preliminaries}
We considered a triple star system in 2+1 hierarchy, namely an eclipsing binary accompanied 
by a third component. The masses of stars in the binary are $m_0$ and $m_1$, and $m_2,$ which 
is that of the third star. We also define $M_1=m_0+m_1,$ the total mass of the binary, and 
$M_2=M_1+m_2,$ the total mass of the system. The configuration of the triple is described 
using the Jacobi system of coordinates, namely $\vv{r}_1,$ describing the position of star~1 
with respect to star~0, and $\vv{r}_2,$ the position of star~2 with respect to the COM of the 
binary. The COM of the whole system is assumed at the origin of the inertial system. The 
vectors $\vv{r}_1$ and $\vv{r}_2$ are represented using Keplerian orbital elements, either 
fixed (Sects.~\ref{liteold} and \ref{litenew1}) or osculating (Sect.~\ref{litenew2}). In 
what follows, we provide results for a general configuration of the triple system using the 
osculating Keplerian elements of both orbits. We only assume that the eccentricity, $e_1,$ 
of the binary orbit is small and introduce non-singular variables $k_1=e_1\cos\omega_1$ 
and $h_1=e_1\sin\omega_1$, where $\omega_1$ denotes the argument of pericenter deduced from 
the tangent plane to the celestial sphere. Note that the generalization to nonzero 
eccentricity, $e_1,$ is straightforward, although algebraically complex. At this moment, we 
postponed this extension of the present analysis to future work. The inclinations $i_1$ of 
the binary orbit and $i_2$ of the third star orbit are arbitrary, the former being only 
constrained by the existence of eclipses.

\subsection{Geometric condition of eclipse mean epoch}\label{ecl}
We adopted a standard frame in which the $XY$ reference plane is tangent to the celestial 
sphere and the $Z$ axis is directed to the observer. The eclipse epochs are defined by 
configurations of minimum sky-plane separation of the stars in the binary. Denoting 
$\vv{r}_1=(x_1,y_1,z_1)^{\rm T}$ in Cartesian coordinates, the epochs of eclipses minimize 
the function $F=x_1^2+y_1^2$. Adopting the elliptic solution for $\vv{r}_1$ and replacing 
the derivative with respect to time, $t,$ by the derivative with respect to the true 
anomaly, $f_1$, the eclipse epochs satisfy 
\begin{equation}
 \frac{d}{dt}\left(x_1^2+y_1^2\right)=n_1\eta_1\left(\frac{a_1}{r_1}\right)^2 
  \frac{d}{df_1}\left(x_1^2+y_1^2\right)=0, \label{EC0}
\end{equation}
 where $n_1$ and $a_1$ denote the mean motion and semimajor axis of the binary orbit and 
 $\eta_1=\sqrt{1-e_1^2}$. It is also customary to introduce the argument of latitude 
 $\phi_1=\omega_1+f_1$ and a related auxiliary angle, $\theta = \phi_1-\theta_\pm$, where 
 $\theta_+=\pi/2$ for primary eclipses and $\theta_-=3\pi/2$ for secondary eclipses.%
\footnote{The primary eclipse occurs when star~1 eclipses star~0, and the secondary
 eclipse occurs when star~0 eclipses star~1. In eclipses, $\theta$ is a small angle that
 justifies the subtraction of $\theta_\pm$ from the argument of latitude in its definition.}
Finally, denoting $x=\sin\theta$, the eclipse epochs satisfy \citep[see, e.g.,][]{gg1983}
\begin{equation}\label{EC1}
 \cB \,x = \left(\cA-x\right) \,\sqrt{1-x^2},
\end{equation}
with
\begin{equation} \label{cAB}
 \cA = \mp \cI\,k_1, \quad
 \cB = \pm \left(\cI+1\right)\, h_1, 
\end{equation}
where
\begin{equation}\label{Idef}
 \cI = \frac{\cos^2 i_1}{\sin^2 i_1}, \quad \cI+1 = \frac{1}{\sin^2 i_1}.
\end{equation}
The upper sign in Eq.~(\ref{cAB}) corresponds to the primary eclipses, while the lower sign 
corresponds to the secondary eclipses. Equation~(\ref{EC1}) has a solution:
\begin{equation}
 x =  \frac{\cA}{1+\cB}+O\left(e_1^4\right), \label{xsol}
\end{equation}
which implies $\theta=\cA +O\left(e_1\right)$ (obviously, $\theta$ or $\phi_1$ are defined 
up to a factor $2\pi k$, where $k$ is an arbitrary integer number). At the epoch of 
eclipses, $\sin\phi_1^\pm=\pm1+O\left(e_1^2\right)$, $\cos\phi_1^\pm=\cI k_1+O\left(
e_1^3\right),$ and at the distance of the two stars in the binary system $r_1^\pm=a_1
\left[1\mp h_1+O\left(e_1^2\right)\right]$.

\subsection{Light-time effect: Traditional description}\label{liteold}
Having set the COM of the whole triple system at the origin of the inertial 
frame, the inertial positions of the three stars are given by
\begin{eqnarray}
 \vv{R}_0 & = & -\mu_2 \vv{r}_2 - \mu_- \vv{r}_1, \label{r0iner} \\
 \vv{R}_1 & = & -\mu_2 \vv{r}_2 + \mu_+ \vv{r}_1, \label{r1iner} \\
 \vv{R}_2 & = & \phantom{-}\left(1-\mu_2\right) \vv{r}_2 , \label{r2iner} 
\end{eqnarray}
where $\mu_-=m_1/M_1$, $\mu_+=m_0/M_1$ and $\mu_2=m_2/M_2$ (with the obvious constraint 
$\mu_-+\mu_+=1$). The inertial position of the COM of the binary is given by the first 
term on the right-hand side of Eqs.~(\ref{r0iner}) and (\ref{r1iner}).

In this work, we chose the $Z=0$ of the global COM as the reference level for LITE.%
\footnote{While this choice is arbitrary, we note a difference compared to the work
 of \citet{conroy2018}, where the reference level for LITE was set at the COM of the binary.}
Denoting parameters of the third-star orbit binary's eclipses, $r_2^\pm$ and $\phi_2^\pm$, 
namely their distance and argument of latitude, the finite speed $c$ of light implies 
a variation of eclipse epochs, $\Delta t_\pm,$ given by
\begin{equation}
 c \Delta t_\pm = \mu_2 r_2^\pm \sin \phi_2^\pm \sin i_2- \mu_\pm a_1
  \left[1\mp h_1+O\left(e_1^2\right)\right]\sin i_1. \label{lite0}
\end{equation}
The first term in Eq.~(\ref{lite0}) arises from the motion of the binary's COM along
the observer's line of sight. This is what most of the literature adopts as the LITE
\citep[e.g.,][]{hinse2012b,borko2015,borko2025}. The second term is due to the motion of
the binary components with respect to their COM. This effect is small and is most often 
neglected. However, it is important to keep it in our formulation, because it combines 
with the inner component of the time correction discussed in Sect.~\ref{litenew1}.

\section{Light-time effect: Fine-tuning}\label{litenew}

\subsection{Light-travel time in the binary system}\label{litenew1}
We now return to the condition set by Eq.~(\ref{EC0}) for eclipse epochs and its solution
outlined in Sect.~\ref{ecl}. The projected coordinates $x_1$ and $y_1$ of the relative vector
between the stars in the binary are given by $x_1=\pm \left(\vv{R}_1-\vv{R}_0\right)\cdot
\vv{e}_X$ and $y_1=\pm \left(\vv{R}_1-\vv{R}_0\right)\cdot \vv{e}_Y$, where $(\vv{e}_X,\vv{e}_Y)$
are unit vectors defining the plane tangent to the celestial sphere. A fundamental implicit 
assumption in Sect.~\ref{ecl} was that both vectors $\vv{R}_0$ and $\vv{R}_1$ are given at
the same time, $t$, which is eventually the epoch of the eclipses used in Eq.~(\ref{EC0}).
Because of the finite speed of light, this setup is not precise and needs to be clarified. 

We now consider the situation of the primary eclipse at time $t$ such that the star indexed $1$ 
is at $\vv{R}_1(t)$. In this case, the star indexed $0$ is seen at some retarded time $t-
\Delta t$, thus $\vv{R}_0(t-\Delta t)=\vv{R}_0(t)-\dot{\vv{R}}_0(t)\Delta t+O(\Delta t^2)$, 
where the overdot mean time derivative. The time interval $\Delta t$, which is small enough
to neglect its higher powers, is set by the light travel from star~0 to star~1 along the 
observer's line of sight; thus, $|(\vv{R}_1(t)-\vv{R}_0(t-\Delta t))\cdot\vv{e}_Z|=
c\Delta t$. Since velocities of stars in the triple system are small with respect to the 
speed of light, we may approximate $c\Delta t = |z_1|$. Returning to the eclipse condition,
the projected $x-$ and $y-$coordinates of the relative vector in (\ref{EC0}) should now 
read $x_1'=x_1+\left(\dot{\vv{R}}_0\cdot \vv{e}_X\right)\Delta t$ and $y_1'=y_1+
\left(\dot{\vv{R}}_0\cdot \vv{e}_Y\right)\Delta t,$ with all coordinates given at a common 
time, $t$. The eclipse now follows from the minimization of $F'={x_1'}^2+{y_1'}^2$ or
\begin{equation}
 F'=F-\left[2\mu_2\left(x_1\dot{x}_2+y_1\dot{y}_2\right)+\mu_-\frac{dF}{dt}\right]
  \frac{|z_1|}{c}. \label{EC2}
\end{equation}
Consider now that the reference eclipse, as given by the approximate solution in 
Sect.~\ref{ecl}, occurred at a certain epoch $t_+$ (i.e., at which $F$ has been minimized). 
The minimization of $F'$ in Eq.~(\ref{EC2}) implies a different but close epoch of $t_++
\Delta t_+$. In what follows, we determine two contributions of $\Delta t_+$ from the two 
terms in the bracket on the right-hand side of Eq.~(\ref{EC2}), starting with the latter --to
be denoted $\Delta t_+^{\rm in}$, since it depends uniquely on motion of the stars in 
the binary-- and following with the former --to be denoted $\Delta t_+^{\rm out}$, since 
it depends on the transverse motion of the binary COM due to the third star.

We recall that the second term in Eq.~(\ref{EC2}) is already a small correction, so we 
can write
\begin{eqnarray}
 \frac{dF'}{dt}\biggl|_{t_++\Delta t_+^{\rm in}}&=&\frac{dF}{dt}\biggl|_{t_++\Delta t_+^{\rm in}}
  -\frac{\mu_-}{c}\frac{d}{dt}\left[|z_1|\frac{dF}{dt}\right]_{t_+}\nonumber \\
 &=& \frac{d^2F}{dt^2}\biggl|_{t_+} \left[\Delta t_+^{\rm in}-\frac{\mu_- |z_1^+|}{c}\right]+
  O\left(\Delta t_+^{{\rm in}\,2}\right)=0, \label{EC3}
\end{eqnarray}
from which it directly follows that $c\Delta t_+^{\rm in}=\mu_- |z_1^+|=\mu_- a_1 \left[1- 
h_1+O\left(e_1^2\right)\right]\sin i_1$. This term combines with the corresponding 
contribution in the second term in the right-hand side of Eq.~(\ref{lite0}) to become 
$c\Delta t_+^{\rm in}=a_1 (\mu_--\mu_+)(1-h_1)\sin i_1$.

A more interesting contribution of $\Delta t_+^{\rm out}$ arises from the first term in the 
right-hand side of Eq.~(\ref{EC2}). Here, we have
\begin{equation}
 \frac{dF'}{dt}\biggl|_{t_++\Delta t_+^{\rm out}}=\frac{d^2F}{dt^2}\biggl|_{t_+} \Delta
  t_+^{\rm out}-\frac{2\mu_2}{c}\frac{d}{dt}\biggl[|z_1| \bigl(x_1\dot{x}_2+y_1\dot{y}_2
  \bigr)\biggr]_{t_+}=0, \label{EC4}
\end{equation}
implying
\begin{equation}
 c\Delta t_+^{\rm out} = 2\mu_2 \frac{d}{dt}\biggl[|z_1 |\bigl(x_1\dot{x}_2+y_1\dot{y}_2
  \bigr)\biggr]_{t_+}\,\left(\frac{d^2F}{dt^2}\biggl|_{t_+}\right)^{-1}. \label{EC5}
\end{equation}
Given the subtleness of the effect, we restricted ourselves to the limit of the circular 
orbit of the binary. In this approximation, we have $d^2 F/dt^2|_{t_\pm} = 2 n_1^2 
a_1^2\sin^2 i_1+O(e_1)$. Note that $d^2 F/dt^2 =2\left(\dot{x}_1^2+\dot{y}_1^2+x_1\ddot{x}_2
+y_1\ddot{y}_2\right)$. If we were to approximate the second derivative of $F$ at eclipse 
epochs by only the velocity terms and neglecting the contribution of the acceleration 
terms, we would have $d^2 F/dt^2|_{t_\pm} \simeq 2 n_1^2 a_1^2$. As a result, the factor 
$\sin i_1$ in the final formula (\ref{EC7}) would change from the denominator to the 
numerator \citep[as in][]{conroy2018}. 

Evaluating the same effects for the secondary eclipses, we obtained the first contribution 
$c\Delta t_-^{\rm in}=\mu_+ |z_1^-|=\mu_+ a_1 \left[1+ h_1+O\left(e_1^2\right)\right]
\sin i_1$, which now adds with the second term in the right-hand side of Eq.~(\ref{lite0}), 
making it $c\Delta t_-^{\rm in}= a_1 (\mu_+-\mu_-)(1+h_1)\sin i_1$. This effect remains 
of limited importance for stellar systems in which the masses $m_0$ and $m_1$ are 
comparable. More significant is the second correction, now reading as
\begin{equation}
 c\Delta t_-^{\rm out} = -2\mu_2 \frac{d}{dt}\biggl[|z_1 |\bigl(x_1\dot{x}_2+y_1\dot{y}_2
  \bigr)\biggr]_{t_-}\,\left(\frac{d^2F}{dt^2}\biggl|_{t_-}\right)^{-1}. \label{EC6}
\end{equation}

Altogether, the resulting inner part of the LITE correction for primary and secondary 
eclipses reads
\begin{equation}
 c\Delta t_\pm^{\rm in}= \mp a_1 (\mu_+ -\mu_-)(1\mp h_1)\sin i_1, \label{lite1}
\end{equation}
and it requires non-equal-mass binaries (or a star--planet system) to be nonzero. Even 
then, (i) its amplitude is small, and (ii) if the eccentricity contribution $\pm h_1$ 
is neglected, it represents merely a constant shift of eclipse timing. However, the 
opposite sign at primary and secondary eclipses produces their asymmetry, which may 
be measurable. This effect is well known, applied in exoplanet transit studies 
\citep[e.g.,][]{fab2010,conroy2018}, and has been discussed in \citet{kap2010} in the 
context of accurate eclipse measurements of binary white dwarfs. In principle, the 
measured shift may provide a useful constraint on binary-star masses, but in the real 
world a degeneracy with eccentricity determination (thus the $\pm h_1$ term) would 
complicate the situation.
\begin{figure*}[t]
 \includegraphics[width=0.95\textwidth]{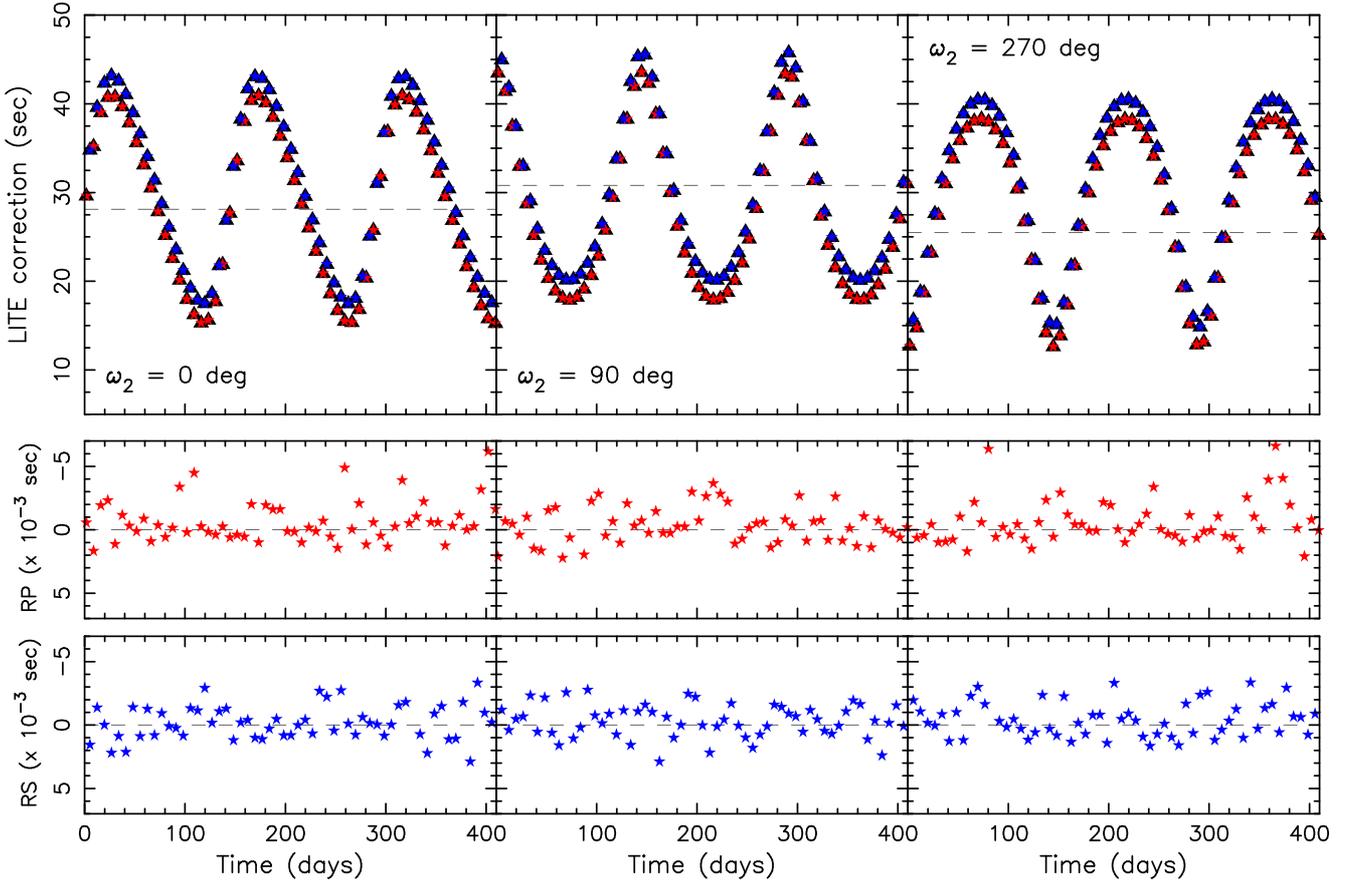}
 \caption{Upper panels: LITE correction due to finite light travel time in the binary system: black triangles
  are from our analytical model (e.g., $\Delta t_\pm^{\rm in}$ and $\Delta t_\pm^{\rm out}$ from
  Eqs.~\ref{EC3} and \ref{EC5} for primary eclipses), red and blue stars from direct numerical
  integration of light travel through the the system (primary and secondary eclipses). We used parameters
  of a well constrained multiple system $\xi$~Tauri \citep[e.g.,][]{nemr2016} that harbors inner triple
  system with a distant fourth component (for sake of simplicity the effect of the fourth star was
  neglected in this example). The triple system is near coplanar with inclination $i\simeq 87^\circ$, inner
  and outer periods $P_1=7.146$~d and $P_2=146$~d, eccentricities $e_1=0$ and $e_2=0.2$. The results are
  shown for three different values of the argument of pericenter of the outer orbit, $\omega_2=0$ (left
  panel), $\omega_2=\pi/2$ (middle panel), and $\omega_2=3\pi/2$ (right panel; the present value is
 $\simeq 10^\circ$ and precesses with a rate $\dot{\omega}_2\simeq 2^\circ$~yr$^{-1}$ due to interaction
  with the eclipsing binary). The masses are $m_0=2.23$~M$_\odot$, $m_1=2$~M$_\odot$ and $m_2=3.74$~M$_\odot$.
  Time at abscissa in days covers three revolutions of the outer orbit, LITE on the ordinate in seconds.
  The mean value of the correction $\simeq a_1/(2c)$ would combine with the second term on the right hand
  side of Eq.~(\ref{lite0}) to be near zero. The $P_2$-periodic variation with an amplitude of $\simeq 15$~s
  is significant and exceeds uncertainty of the TESS photometry. {Middle and lower panels: the
  residuals between the exact numerical simulation and our analytic formulation are shown for primary
  eclipses by red stars in the middle panel and secondary eclipses by blue stars in the lower panel. Their
  average value is zero and no signal above $\simeq 0.005$ seconds (namely $0.03$\% of the LITE correction)
  is observed}.}
 \label{f1}
\end{figure*}

The more significant outer component of the LITE corrections of (\ref{EC5}) and (\ref{EC6}) for the 
primary and secondary eclipses may be expressed using the Keplerian orbital elements of both the binary 
and the outer orbit and given in a compact form:
\begin{equation}
 c\Delta t_\pm^{\rm out} = \frac{\mu_2 a_2}{\eta_2\sin i_1} \, \cZ_\pm, \label{EC7}
\end{equation}
with
\begin{eqnarray}
 \cZ_\pm &\!\!\!=\!\!\!& \sin\left(\phi_2^\pm+\Delta\Omega\right) + e_2\sin\left(\omega_2+\Delta
  \Omega\right)- \nonumber \\ 
 &\!\!\!\phantom{=}\!\!\! & 2\,\left(\cos\phi_2^\pm+e_2\cos\omega_2\right) \sin^2\frac{i_2}{2}
  \sin\Delta\Omega -\nonumber \\ 
 &\!\!\!\phantom{=}\!\!\! & \eta_2\cos i_1\, \frac{n_2}{n_1}\left(\frac{a_2}{r_2^\pm}\right)^2
  \biggl[\sin\left(\phi_2^\pm+\Delta\Omega\right)-
 \nonumber \\  &\!\!\!\phantom{=}\!\!\! & \hspace*{27mm}
 2\sin\phi_2^\pm \sin^2\frac{i_2}{2} \cos\Delta\Omega \biggr]; \label{EC8}
\end{eqnarray}
$\Delta\Omega=\Omega_2-\Omega_1$ is the difference in nodal longitudes of the two orbits. In the case of
a coplanar configuration of the triple system, $i_2=i_1=i$ and $\Delta\Omega=0$, Eq.~(\ref{EC8}) 
takes a simpler form:%
\footnote{In the case of a coplanar system with retrograde motion of the third component, one has
 $\Delta \Omega=\pi$ and $i_2=\pi-i_1$. The LITE correction $c\Delta t_\pm$ is still given by
 Eq.~(\ref{EC9}), but now (i) with a reversed sign, and (ii) $i=i_1$ on the right-hand side specifically.}
\begin{eqnarray}
 c\Delta t_\pm^{\rm out} &=& \frac{\mu_2 a_2}{\eta_2\sin i} \frac{n_2}{n_1}\biggl[\sin\phi_2^\pm+
  e_2\sin\omega_2-\nonumber \\ & & \hspace*{17mm} \eta_2\cos^2 i\,\frac{n_2}{n_1}\left(\frac{a_2}{r_2^\pm}
 \right)^2\sin\phi_2^\pm\biggr]. \label{EC9}
\end{eqnarray}
Note that the general formula (\ref{EC8}) is also valid for $i_2=\pi/2$ when both the node $\Omega_2$ and
the argument of latitude $\phi_2^\pm$ of the outer orbit are undefined. However, the sum
$\Delta\Omega+\phi_2^\pm$ and $\Delta\Omega+\omega_2$ are well-defined and continuous.

The last term in the bracket of Eq. (\ref{EC8}),{ which is new in the analytical form,} is typically small,
but the first term represents a contribution to LITE shift, whose magnitude is only a factor $n_2/n_1$
smaller than the principal effect in Eq. (\ref{lite0}). In systems near the stability limit for which the
period ratio $P_2/P_1$ is not too large (e.g., $\leq 20),$  the correction in Eq. (\ref{EC8}) or
(\ref{EC9}) for coplanar systems may amount to $5-15$\% of Eq. (\ref{lite0}). Also note that the effect
is maximum for coplanar geometry \citep{conroy2018}, but it is easy to verify using Eq. (\ref{EC8}) that
non-coplanar geometries may not compromise the LITE correction due to the finite light-travel time in the
binary system. In fact, even for the case of the outer orbit perpendicular to the orbital plane of the binary,
that correction may be exactly the same as for the coplanar geometry depending on the precise architecture of
orbits in the triple system.{ Finally, we point out that our more rigorous analysis considering
elliptic motion of both the binary and the third star in the system allowed us to find that the factor
$\sin i_1$ in Eqs.~(\ref{EC7}) and (\ref{EC9}) should stand in as a denominator of the multiplicative factor
rather than its numerator. This is particularly important for systems where the binary inclination is not
too close to $90^\circ$}.

In order to illustrate relevance of the LITE correction due to the finite travel time of light in the
binary system, and also to test its accuracy, in Fig.~\ref{f1} we show results for the well-constrained
multiple system $\xi$~Tauri \citep[see, e.g.,][]{nemr2016}. For the sake of simplicity, we neglected effects
of a distant fourth component in the system at this moment and only considered the compact triple core, which
consists of
an eclipsing binary with a $P_1=7.146$~d period, and a third component with $P_2=146$~d revolution about the
COM of the binary. The system is nearly coplanar, with an inclination of $i\simeq 87^\circ$. The binary has
a nearly circular orbit, and the outer orbit has an eccentricity of $e_2\simeq 0.2$. The best estimates of
the three masses are $m_0=2.23$~M$_\odot$, $m_1=2$~M$_\odot,$ and $m_2=3.74$~M$_\odot$. The three upper panels
of Fig.~\ref{f1} show results for three different values on the argument of pericenter, $\omega_2,$ of the
outer orbit. We compare an analytic formulation of LITE correction present in this section (black triangles)
with results of a fully fledged numerical analysis that includes effects of light propagation in the system
(red and blue stars for primary and secondary eclipses). There is a fairly good agreement of the two
results, with the analytic formulation being much less demanding in terms of CPU. The amplitude of the principal LITE
component from Eq. (\ref{lite0}) is about five minutes, while the amplitude of Eq. (\ref{EC7}) is about $15$
seconds. In passing, we also note that the eclipse shifts due to the inner effect of Eq. (\ref{lite1}) are
$\simeq 3.2$~s for $\xi$~Tau. Given the formal accuracy of the eclipse timing from TESS photometry,
$\lesssim 5$ seconds in the latest sectors, both contributions are relevant. 

{To further probe the accuracy of the analytic formulation, we show, in the middle and lower
panels of Fig.~\ref{f1}, residuals between the numerical simulation and the analytical model. 
Satisfactorily, their mean value is zero, and no systematic residual signal exceeds $\simeq 0.005$ seconds
\citep[amounting to $\simeq 0.03$\% of the LITE correction amplitude, apparently
improving accuracy of a similar simulation reported in][Fig.~11, by more than one order of magnitude]{conroy2018}.
In fact, the residuals are at the level of numerical precision of our simulation.%
\footnote{In passing, we note that the simulation shown in Fig.~\ref{f1} did not account for the observed
 periastron precession, $\dot{\omega}_2\simeq 2^\circ$~yr$^{-1}$, of the outer orbit. In the case of coplanar
 configuration, this effect would contribute via an additional LITE correction:
\begin{equation}
 c\Delta t_\pm^{\rm out} = \frac{\mu_2 a_2}{\sin i} \frac{\dot{\omega}_2}{n_1}\biggl[\frac{r_2^\pm}{a_2}\sin\phi_2^\pm
  -\frac{\cos^2 i}{\eta_2}\,\frac{n_2}{n_1}\left(\sin\phi_2^\pm+e_2\sin\omega_2
 \right)\biggr], \label{EC9bis}
\end{equation}
 with an amplitude of about $0.05$~s for the $\xi$~Tauri system. When included, the residuals drop to a $0.005$~s
 level, as in Fig.~\ref{f1}.}
However, it should be
pointed out that the optimum accuracy hinges on the assumption of binary's circular orbit ($e_1=0$).
Since the corrections of $\Delta t^{\rm out}_\pm$ in Eqs.~(\ref{EC7}) to (\ref{EC9}) do not contain 
$e_1$-dependent
terms, accuracy would degrade with increasing $e_1$ value. A generalization of our results that would
include $e_1$-dependent terms is left for future work (our tests show that the loss of accuracy scales 
linearly with increasing $e_1$ values for low enough values)}.

\subsubsection{Systemic velocity and higher multiplicity}
The formulation given in the previous section assumed an isolated triple system. Here, we briefly discuss 
the effect of systemic velocity of the triple and the situation when the triple is embedded in a system 
of higher multiplicity (i.e., accompanied with other, more distant stars on bound orbits).

We next assume that the COM of the triple system is given by ${\bf R}(t)$, such that at an arbitrary epoch 
of $t=0$ coincides with the origin of an inertial system, and we denote ${\bf V}(t)=d{\bf R}/dt$ its 
velocity. The previous formulation of LITE has to be extended by an obvious component,
\begin{equation}
 c\Delta t_\pm = - R_Z(t), \label{lite2}
\end{equation}
in Eq.~(\ref{lite0}), and the correction's outer term,
\begin{equation}
 c\Delta t_\pm^{\rm out} = \mp 2 \frac{d}{dt}\biggl[|z_1 |\bigl(x_1\dot{R}_X+y_1\dot{R}_Y\bigr)
  \biggr]_{t_\pm}\,\left(\frac{d^2F}{dt^2}\biggl|_{t_\pm}\right)^{-1}, \label{EC10}
\end{equation}
in Eqs.~(\ref{EC5}) and (\ref{EC6}). The constant systemic velocity of ${\bf V}={\bf V}_\perp + 
V_\parallel\,{\bf e}_Z$, where ${\bf V}_\perp$ is tangent to the sky plane of the triple system 
provides
\begin{equation}
 c\Delta t_\pm^{\rm out} = \frac{V_\perp}{n_1}\,\frac{\cos\Delta\Omega}{\sin i_1},\label{lite3}
\end{equation}
where $\Delta \Omega$ is the angular distance between ${\bf V}_\perp$ and the ascending node of the 
binary. To illustrate the effect, we considered the $\xi$~Tau system from Fig.~\ref{f1}, where 
$V_\parallel \simeq 8.8$ km~s$^{-1}$, $V_\perp \simeq 2.0$ km~s$^{-1}$, $\Delta\Omega\simeq 
160^\circ,$ and $n_1a_1\simeq 180$ km~s$^{-1}$ \citep[e.g.,][]{nemr2016}. The radial component 
of Eq.~(\ref{lite2}) accumulates to $\sim 2.6$~hr in a decade, but it is merely a constant 
drift, which is routinely included in data analysis. The transverse part of Eq.~(\ref{lite3}) 
amounts to only an $\simeq 0.6$~s retardation of eclipses, and it may thus be neglected with 
currently available observations. This small effect is due to the anomalously small sky-plane 
velocity $V_\perp$ of the system. \citet{conroy2018} analyzed the Gaia Data Release~1 data and 
found that the distribution of $V_\perp\cos\Omega$ has one and two sigma values of $\simeq 29$ 
km~s$^{-1}$ and $\simeq 58$ km~s$^{-1}$. Therefore, a time shift in eclipses of up to $10$~s 
is possible. In such a case, if neglected, it could erroneously mis-manifest in the $e_1$ 
solution at the $\simeq 0.01$ level.

\subsection{Coupling with dynamical interaction}\label{litenew2}
Another small correction to the traditional formulation of the LITE arises from the dynamical 
interaction between the binary and the orbit of the third star in the system. This is because 
LITE, as described in Sect.~\ref{liteold}, assumes simple elliptical orbits for the stars in 
the binary described by $\vv{r}_1$ and the third star about their COM described by $\vv{r}_2$. 
In reality, both trajectories are subject to a mutual gravitational interaction in the system.

In this work, we determined the principal correction to LITE due to mutual interaction of 
stars in the system, leaving a more detailed formulation to future studies. From the multitude 
of interaction contributions, we considered those with a $P_2$ periodicity of the outer orbit.%
\footnote{The secular effects, such as periastron precession or inclination variation, may be
 trivially included in the previous formulation, allowing time dependence of the relevant
 orbital elements (as an example, see footnote~5).}
We applied a first-order perturbation approach to evaluate the leading correction to the 
binary's COM position in the $e_1=0$ and $e_2=0$ limits and assumed a coplanar 
configuration. The principal LITE contribution in Eq. (\ref{lite0}) reads $c\Delta t = -Z$, 
where $Z=-\mu_2 r_2 \sin\phi_2 \sin i$ is the position of the binary's COM along the line 
of sight of the observer. By denoting its perturbation by $\Delta Z$, we directly obtained 
the LITE correction as $c\Delta t^{\rm c} = -\Delta Z$. We used results from \citet{bv2015}, 
the authors of which constructed a Hamiltonian describing secular dynamics of the triple 
system using Lie--Hori canonical transformation algorithm of short-period term elimination. 
The associated generating function, $S,$ was useful for our task, as
\begin{equation}
 \Delta Z = -\left\{Z;S\right\}. \label{dz}
\end{equation}
To keep the work in the lowest order, we also chose the generator, $S,$ describing the 
quadrupole coupling of the binary and outer orbits in the triple system and restricted it 
to its part independent of the binary's eccentricity, $e_1$. In this approximation, we have
\begin{equation}
 S=-\frac{2}{3}\frac{C_2}{n_2\eta_2^3}\left(f_2-\ell_2+e_2\sin f_2\right)
 , \label{genS}
\end{equation}
where 
\begin{equation}
 C_2=\frac{3}{8}\frac{G\, m_2\, m_1'}{a_2}\left(\frac{a_1}{a_2}\right)^2
 , \label{genS1}
\end{equation}
$m_1'=m_0m_1/M_1$ is the reduced mass of the binary, and $\ell_2$ is the mean anomaly of 
the outer orbit. We used Delaunay elements to compute the Poisson bracket $\left\{Z;S\right\}$ 
in Eq.~(\ref{dz}), obtaining a small correction of LITE:
\begin{equation}
 c\Delta t_\pm^{\rm c} =\mu_2 a_2\sin i\,\Lambda\,\left(\sin\phi_2^\pm - \frac{1}{6}
  \cos\phi_2^\pm\right), \label{litez}
\end{equation}
where
\begin{equation}
 \Lambda = \frac{3}{4}\frac{n_2}{n_1}\frac{m_1'}{M_2}\sqrt{\frac{M_2 a_1}{M_1 a_2}},
  \label{lite3bis}
\end{equation}
and in a particular case of equal-mass stars in the binary, we have
\begin{equation}
 \Lambda = \frac{3}{16}\left(\frac{n_2}{n_1}\right)^{4/3}\left(\frac{M_1}{M_2}
  \right)^{2/3} ,\label{lite3bbis}
\end{equation}
or $\Lambda \simeq 0.013\,(M_1/M_2)^{2/3}$ for systems near the stability limit period 
ratio $P_2/P_1\simeq 7.5$. For example, the amplitude of LITE correction in 
Eq.~(\ref{litez}) amounts to $\simeq 0.7$~s for the $\xi$~Tau system ($\Lambda\simeq 2.2
\times 10^{-3}$).
As a result, the effect is truly small, and its relevance is yet to be proven.

\section{Conclusions}\label{concl}
We developed a new take on the previously published concept of asymmetric transverse 
velocities of components in a binary for time shifts in exact epochs of their primary 
and secondary eclipses \citep{conroy2018}. The finite travel time of light in the binary 
system itself is the essence of the effect in both \citet{conroy2018} and our work; 
therefore, both works extend the classical LITE known in stellar astronomy for a 
century. We provide a simple analytical formulation valid for generally non-coplanar 
triple systems containing an eclipsing binary. Although small, the correction to 
the classical LITE is fully relevant in many systems, with precise photometry provided 
by space-born surveys available. In Sect.~\ref{litenew1}, we use the multiple stellar 
system $\xi$~Tauri and an exemplary case. The strength of our result is in its 
simplicity, which may help to preconstrain model parameters prior to a fully fledged 
numerical model being applied.

We also analyzed the coupling of LITE with long-period dynamical perturbation of the 
orbits in the triple system and provided estimates of its contribution to the eclipse 
timing variation. The corresponding term appears small, at a maximum $0.5-1$\% of 
the classical LITE. Nevertheless, advancements in technology may prove to be relevant 
in the future.

\begin{acknowledgements}
 The author thanks Petr Zasche and Marek Wolf for useful information on the topic of
 this paper, and the referee whose suggestions helped to improve its final version.
 Special thanks belong to S\l{}awek Breiter for his help in computing the Poisson bracket
 (\ref{dz}). The support of the Czech Science Foundation is acknowledged (grant 25-16507S).
 This work is dedicated to the memory of Dr.~Pavel Mayer,
 one of the pioneers in rigorously discerning the role of the
 light time effect in observations of eclipsing binaries.
\end{acknowledgements}

\bibliographystyle{aa}

\end{document}